information system built from those products using sound system and security engineering principle are sufficiently trustworthy.

**References**

1. **National** Institute of Standard and Technology Special Publication 800-53, Recommendation Security controls for Federal Information Systems, February 2005.

2. **National** Institute of Standard and Technology Special Publication 800-53 Revision 4, Security and Privacy Controls for Federal Information Systems and Organizations, February 2013.

*Статья представлена научным руководителем д-ром техн. наук, профессором М. В. Буйневичем.*

**УДК 004.72; 004.715**

**Тиамийу А. Осуолале**

# К ВОПРОСУ О МОДЕЛИРОВАНИИ МЕХАНИЗМА ДОВЕРЕННОЙ МАРШРУТИЗАЦИИ

*Среди механизмов защиты данных в компьютерных сетях рассмотрена доверенная маршрутизация. Выбран способ ее моделирования и обоснован сетевой симулятор.*

*механизмы защиты, доверенная маршрутизация, моделирование, сетевой симулятор.*

Известны и достаточно широко применимы различные механизмы защиты данных в IP-сетях, такие как: шифрование (криптографическое закрытие данных); авторизация; изоляция компьютерных сетей (фильтрация трафика, скрытие внутренней структуры и адресации, противодействие атакам на внутренние ресурсы etc); выявление и нейтрализация действий компьютерных вирусов и др. [1].

Среди таких механизмов выделим доверенную маршрутизацию (ДМ), которая является процессом планирования маршрутов через подмножество, так называемых, доверенных узлов, а также организации передачи данных по таким маршрутам от источника к месту назначения, исключая возможности подмены, модификации или внедрения какой-либо другой информации в потоки данных [2].

Собственно, механизм ДМ известен достаточно давно (и 10 лет как стандартизован), однако крайне редко (и только в экспериментальных или узкоспециализированных сетях) применяется на практике. Причина такого положения, в частности, кроется в том, что ДМ еще недостаточно глубоко исследована научным сообществом и поэтому у потенциальных сетевых операторов нет требуемых знаний ее положительных свойств и ограничений.

В этой связи актуальным является всестороннее исследование механизма ДМ всеми доступными средствами на предмет гарантированно безопасной и надежной передачи трафика. Главным средством здесь выступает моделирование, к которому традиционно предъявляются следующие требования [3]:

– адекватность, т. е. соответствие модели исходной реальной системе и учет, прежде всего, наиболее важных качеств, связей и характеристик;

– точность, т. е. степень совпадения полученных в процессе моделирования результатов с заранее установленными (желаемыми);

– экономическая целесообразность, т. е. точность получаемых результатов и общность решения задачи должны увязываться с затратами на моделирование.

Кроме того, моделирование должно оптимизировать уровень детализации, поскольку «крупное» моделирование (допущения) дает слишком грубые результаты, а «мелкое» моделирование (много деталей) приведет к очень длительному моделированию, результаты которого, к тому же, являются сложными для анализа.

Важнейшей задачей, доминантно обеспечивающей выполнение перечисленных требований, является выбор и обоснование метода моделирования, например, для исследования механизма ДМ.

Натурное моделирование – отличительной чертой этого метода является подобие моделей реальным системам (они материальны), а отличие состоит в размерах, числе и материале элементов и т.п. Физическая модель позволяет охватить явление или процесс во всём их многообразии, наиболее адекватна и точна, но достаточно дорога, трудоёмка и менее универсальна.

Аналитическое моделирование позволяет исследовать одни физические явления или математические выражения посредством изучения других физических явлений, имеющих аналогичные математические модели. Аналитическая модель удобна в работе, но зачастую имеет слишком принципиальные допущения и ограничения.

Имитационное моделирование – метод, позволяющий строить модели, описывающие процессы так, как они проходили бы в действительности. Здесь главный вопрос состоит в отборе действительно важных факторов для моделирования, так как запрограммировать теоретически «можно все». Такую возможность дают как «классические» языки, так и специализированные – например, GPSS. Лучшие практики (best practices) имитационного моделирования телекоммуникационных сетей и протекающих в них процессов нашли свое отражение в специализированных программных пакетах – сетевых симуляторах.

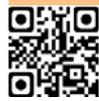

Методы и системы защиты информации, информационная безопасность
в системах связи и телекоммуникаций

Среди многочисленных сетевых симуляторов наиболее популярными являются ns-2, ns-3, OPNET, OMNeT++, QualNet, AnyLogic, GTNetS, и NetSim. Они могут быть условно разделены на категории или сравнены на основании простоты использования, масштабируемости, скорости выполнения, наличия библиотек, уровня поддержки и т. д. И с этих позиций выбор того или иного симулятора определяется, скорее всего, «вкусовыми» предпочтения исследователя (доступность, дружественность интерфейса, навыки работы). Однако с позиций исследования механизма ДМ определяющим будет наличие возможностей моделирования стека дистанционно-векторных, междоменных и внутридоменных протоколов маршрутизации (RIP, IGRP, BGP, EIGRP, OSPF, IS-IS), а также MPLS [4]. Кроме динамического построения маршрутов и использования защищенной сети для передачи данных между отправителем и получателем, модель должна имитировать контроль прохождение потока данных и удаленное управление роутерами.

Наиболее полно этим требованиям отвечает разработанное корпорацией MIL3 средство проектирования вычислительных сетей OPNET (Optimum Network Performance), а конкретно – OPNET Modeler (средство моделирования и анализа производительности сетей, компьютерных систем, приложений и распределенных систем). OPNET Modeler сочетает в себе следующие функциональные возможности по моделированию механизма ДМ: создание иллюстративной схемы защищенной сети с наложением на карту, верификация соединений на предмет соответствия выбранных типов связи соответствующим портам подключенного к ним оборудования, моделирование процесса обработки (теория очередей) и стоимости использования оборудования передачи данных, импортирование существующей конфигурации сети. Также имеется множество библиотек компонентов, в том числе конкретных сетевых устройств, типов соединений, протоколов и сетевых приложений. Существует возможность подключения новых библиотек и создания собственных экземпляров компонентов [5].

Таким образом, актуальной является задача имитационного моделирования механизма доверенной маршрутизации в программной среде OPNET.

**Список используемых источников**

Методы и системы защиты информации, информационная безопасность
в системах связи и телекоммуникаций

*Статья представлена научным руководителем д-ром техн. наук, профессором М. В. Буйневичем.*